\documentclass{article}
\usepackage{spconf,amsmath,graphicx,amsthm,amssymb,pifont,url}
\usepackage{cite} 
\usepackage{multirow}
\usepackage{algorithm}
\usepackage{algpseudocode}
\usepackage{colortbl}  
\usepackage{tabularx}
\usepackage[table,xcdraw]{xcolor}
\usepackage{booktabs}
\usepackage{subcaption}
\usepackage{multirow}
\usepackage[normalem]{ulem}
\newcommand{\cmark}{\ding{51}}%
\newcommand{\xmark}{\ding{55}}%

\title{Unsupervised Noise adaptation using Data Simulation}
\name{Chen Chen$^1$, Yuchen Hu$^1$, Heqing Zou$^1$, Linhui Sun$^2$, Eng Siong Chng$^1$
\sthanks{This research is supported by the Ministry of Education, Singapore, under its Academic Research Fund Tier 2 (MOE2019-T2-1-084).}}
\address{$^1$School of Computer Science and Engineering, Nanyang Technological University, Singapore \\ $^2$ Nanjing University of Posts and Telecommunications, China}
\begin{document}
%
\maketitle
\begin{abstract}
Deep neural network based speech enhancement approaches aim to learn a noisy-to-clean transformation using a supervised learning paradigm. However, such a trained-well transformation is vulnerable to unseen noises that are not included in training set. In this work, we focus on the unsupervised noise adaptation problem in speech enhancement, where the ground truth of target domain data is completely unavailable. Specifically, we propose a generative adversarial network based method to efficiently learn a converse clean-to-noisy transformation using a few minutes of unpaired target domain data. Then this transformation is utilized to generate sufficient simulated data for domain adaptation of the enhancement model. Experimental results show that our method effectively mitigates the domain mismatch between training and test sets, and surpasses the best baseline by a large margin. 
\end{abstract}
\section{Introduction}
\label{sec:intro}
Recent advances of deep learning have brought remarkable progress to speech enhancement technique~\cite{lu2013speech,wang2018supervised,pascual2017segan}. Generally, diverse deep neural networks are designed to convert the noisy speech input to enhanced signal, where the parallel clean speech is served as ground truth to provide supervised information~\cite{valentini2016investigating, xu2014regression}. However, such a data-driven learning paradigm suffers from the mismatch between training and test data distributions: we usually prepare various types of noise in the training set for generalization of enhancement, while the noise types in test set are not always included in the training set~\cite{lin2021unsupervised}. These unseen noises are not applicable for the trained model that only learns the noisy-to-clean transformation from training data, thus resulting in limited enhancement performance~\cite{hou2021learning}. \par
This training-testing difference is generally called domain mismatch in speech enhancement. To address this issue, unsupervised domain adaptation techniques have been widely introduced to adjust the SE model to unseen noise distribution~\cite{ganin2015unsupervised}. It is noted that the “unsupervised” denotes that labels of target domain data are completely unavailable. Mainstream unsupervised noise adaptation methods are summarized into 2 categories: learning the alignment of domain-invariant features~\cite{sun2016deep, morerio2017minimal} and adversarial training~\cite{ganin2016domain,liao2018noise}, where a discriminator is employed as a domain classifier. Despite of effectiveness, we argue that such methods are limited in exploring more general representations while failing to efficiently utilize the relationship between source and target domain. \par
In this paper, we propose a data simulation-based method (UNA-GAN) to handle unsupervised noise adaptation. Different from typical adversarial training, the UNA-GAN aims to learn a clean-to-noisy transformation that directly converts clean speech to noisy speech in target domain. Since clean signal has negligible domain shift\cite{lin2021unsupervised}, it precisely serves as ground-truth signal to supervise the generated noisy speech with a similar distribution of target domain speech. When such a transformation is trained well, a large amount of simulated parallel dataset is available to finetune the SE model, which adapts it to target domain data. \par
The main superiority of UNA-GAN is summarized as follows: (1) High data efficiency. Only several minutes of unlabeled target domain noisy data is required to learn noise distribution in spectrogram, which is simply implemented in practical conditions. (2) Unpaired training examples. The clean-to-noisy transformation can be acquired by the training pairs with mismatched utterances, as the target noise is viewed as primary simulation objective. The intensive experiments demonstrate that the proposed UNA-GAN is able to simulate near-authentic noisy speech and achieves effective noise adaptation to target domain. Furthermore, UNA-GAN surpasses other unsupervised domain adaptation baselines by a large margin in terms of evaluation metrics, even in face of large domain mismatch and low SNR conditions.

\begin{figure*}[t]
\centering
\includegraphics[width=0.95\textwidth]{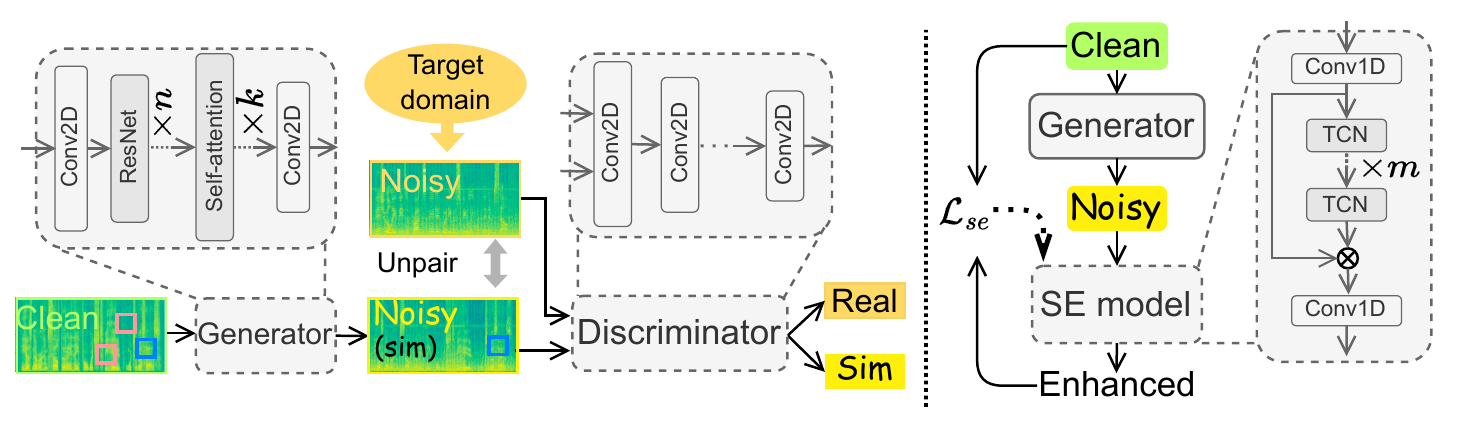}
\vspace{-0.1cm}
\caption{The overview of UNA-GAN. The module with a dashed box is trainable, while that with a solid box is fixed.}
\vspace{-0.2cm}
\label{f1}
\end{figure*}

\section{UNA-GAN Method}
In this section, we first illustrate the research problem of unsupervised noise adaptation and context notations. Then we introduce the proposed UNA-GAN method that consists of a data simulation part and a model adaptation part. The overview structure of UNA-GAN is shown in Fig.~\ref{f1}.

\subsection{Problem Setting}
Consider a source domain $S$ with paired noisy-clean data $(X^S,Y^S)=\{(x_i^S,y_i^S)\}_{i=1}^{N_S}$, where $x_i^s$ and $y_i^s$ respectively denote the noisy speech from source domain and its corresponding ground-truth clean speech. Unsupervised noise adaptation assumes that some noisy data $X^T=\{x_i^T\}_{i=1}^{N_T}$ from another target domain $T$ is unlabeled. Our goal is to find a noisy-to-clean transformation $F_{se}^T$ for target domain data that predict the clean speech label $\{y_i\}_{i=1}^{N_T}$, based on the knowledge $F_{se}^S$ learned from source domain. \par  
We denote the amount of source domain and target domain as $N_S$ and $N_T$. In practice, $N_S$ is much larger than $N_T$, and the speech examples of two domains are unpaired. Due to domain mismatch, the performance of learned $F_{se}^T$ would dramatically degrade when predicting $Y^T$ from $X^T$.   

\subsection{Data Simulation for Target Domain}
Given noisy speech $X^T$ from target domain, the objective of data simulation is to learn a domain transformation $F^{S\sim T}_{gen}$ that mimics the distribution of $X^T$ from clean speech $Y^S$. To this end, we employ a GAN-based structure, as shown in Fig~\ref{f1}, where the training examples $\{Y^S_i$, $X^T_i\}_{i=1}^{N_T}$ are limited and not required to be paired. \par 

\noindent \textbf{Generator and Discriminator}.
Generator $G$ is designed to map the clean magnitude to simulated noisy magnitude. As shown in Fig.~\ref{f1}, it contains symmetrical 2-D convolutional layers with kernel 3$\times $3 for down-sampling and up-sampling, respectively. Among them, we repeat an ResNet block~\cite{he2016deep} for $n$ times to learn deep representations, and each block consists of two convolutional layers with the kernel size of 3$\times$3 followed by one dropout layer. Then we repeat a self-attention layer~\cite{chen2021time} for $k$ times, which is designed to catch global information of utterance. In general, that simulator aims to integrate target noisy features into clean magnitude without any change in shape. \par
Discriminator $D$ is employed to distinguish where the input magnitude come from (\textit{i.e.}, simulated or real). We repeat five 2-D convolutional layers with the kernel size of 4$\times$4 followed by the LeakyReLU activation function. For down-sampling, the stride takes 2$\times$2 for the first three convolutional layers and 1$\times$1 for the last two convolutional layers. During training, the adversarial loss \cite{goodfellow2020generative} is defined as:
\begin{small}
\begin{equation}
\mathcal{L}_{gan} (G,D,X,Y) = \mathbb{E}_{x\sim X^T}\log D(x) +\mathbb{E}_{y\sim Y^S}\log (1-D(G(y)))
\end{equation}
\end{small}
By minimizing this loss, simulated noisy magnitude learns to be visually like the real noisy magnitude of target domain. In this imitation process, the human speech is viewed as invalid information, as the discriminator mainly distinguishes magnitudes in terms of the distribution of background noises. \par
\noindent \textbf{Contrastive learning}. In order to confuse the discriminator, the generator is diligently learned to incorporate similar noise into clean magnitude. However, it might result in over-simulation that overwrites too much useful content. To address it, we employ contrastive learning to maximize the mutual information~\cite{chen2022interactive} between clean magnitude and simulated noisy magnitude as they are paired.\par
As shown in Fig.~\ref{f1}, we first sample 256 patches $\{\hat{z_i}\}_{i=1}^{256}$ in simulated noisy magnitude as query and select its corresponding patches $z$ in clean magnitude. The corresponding pair of patches in the same places are viewed as positive examples ($\hat{z}^i,z^i$), while other mismatch pairs ($\hat{z}^i,z^j$) are viewed as negative examples. Such selected patches are reshaped via two linear layers with 256 units followed by the ReLU activation~\cite{chen2022noise}. Finally, we calculate the cross-entropy loss using the positive and negative training examples as follows:  
\begin{small}
\begin{equation}
\label{lossMPC}
\mathcal{L}_{cl}(G,Y) = \sum_{i=1}^{I} -\log\left[ \frac{e^{(\hat{z}^i \cdot z^i / \tau)}}{e^{(\hat{z}^i \cdot z^i / \tau)} + \sum_{j=1}^J e^{(\hat{z}^i \cdot z^{j} / \tau)} } \right ]
\end{equation}
\end{small}
Besides the input magnitude, we repeat the patch-sampling operation in feature layers for contrastive learning. To this end, the generator is reused that takes simulated noisy magnitude as input. We select further 4 layers in generator, which are the two down-sampling convolutional layers, and the first and the medium residual block. Consequently, the total loss function $\mathcal{L}_{una}$ can be formulated as:   
\begin{small}
\begin{equation}
\mathcal{L}_{una} = \mathcal{L}_{gan}(G,D,X,Y)  + \alpha \mathcal{L}_{cl}(G,Y) + \beta \mathcal{L}_{cl}(G,X) 
\label{loss_total}
\end{equation}
\end{small}
where $\mathcal{L}_{cl}(G,X)$ calculates the same contrastive loss based on noisy data of target domain, which is incorporated to prevent the generator from making unnecessary changes~\cite{park2020contrastive}.

\subsection{Model Adaptation}
We now introduce the mask-based SE model and adaptation strategy using UNA-GAN. \par
\noindent \textbf{Model structure}. As shown in Fig.~\ref{f1}, the SE model first employ a 1-D convolutional layer as encoder, which covert time domain signal to hidden features. Then we employ TCN blocks with same structure of Conv-TasNet~\cite{luo2019conv}, which consists of dilatation convolutional layer~\cite{chen2021time} and two 1-D convolutional layers. In order to expand the receptive field, we repeat the TCN blocks for $m$ times with increasing dilation factors 2$^{m-1}$. The output of final TCN blocks serves as mask that multiplies the output of encoder, which is expected to remove the noise to obtain enhanced feature. Finally, the enhanced feature is converted back to time domain signal by a 1-D convolutional decoder. \par     
\noindent \textbf{Adaptation strategy}. We first train a baseline SE model $F_{se}^S$ with dataset from source domain $\{X_i^{S},Y_i^S\}_{i=1}^{N_S}$. To this end, a multi-scale scale invariant signal-to-distortion ratio (SI-SDR)~\cite{luo2019conv} loss is calculated as following: 

\begin{equation}
\mathcal{L}_{se}= 10\log_{10}(\frac{\|\frac{\left \langle \hat{s},s\right \rangle}{\left \langle s,s\right \rangle} s\|^2}{\|\frac{\left \langle \hat{s},s\right \rangle}{\left \langle s,s\right \rangle} s-\hat{s}\|^2})
\end{equation}
where the $\hat{s}$ and $s$ respectively stand for enhanced time domain signal and clean ground-truth.\par
Given limited $\{X_i^T\}_{i=1}^{N_T}$ from target domain, the same amount of clean speech $\{Y_i^S\}_{i=1}^{N_T}$ are randomly sampled from $\{Y_i^S\}_{i=1}^{N_S}$. Then we train the UNA-GAN with unpaired dataset $\{Y^S_i$, $X^T_i\}_{i=1}^{N_T}$ using $\mathcal{L}_{una}$ in Eq.~(\ref{loss_total}).  After training, the trained-well generator is utilized as domain converter $F^{S\sim T}_{gen}$ from $Y^S$ to $X^T$. Since the data amount of clean speech $N^S$ is usually abundant, a large dataset can be simulated with paired $\{X_i^{T_{sim}},Y_i^S\}_{i=1}^{N_S}$. It is noted that the domain shift of clean speech is negligible, therefore, $\{X_i^{T_{sim}},Y_i^S\}_{i=1}^{N_S}$ can be approximately viewed as $\{X_i^{T_{sim}},Y_i^T\}_{i=1}^{N_S}$, which is subsequently utilized to finetune SE model using $\mathcal{L}_{se}$. \par 

\section{Experiment}
\subsection{Dataset}
We evaluate our method on two datasets: Voice Bank-DEMAND~\cite{valentini2016investigating} and TIMIT~\cite{garofolo1988getting}. Our approach does not rely on additional data information such as domain labels. \par
\noindent \textbf{VoiceBank} The training set (source domain) contains $N_S=11572$ noisy utterances from 28 speakers and is mixed by 10 different types with four SNR levels (0, 5, 10, and 15 dB) at a sampling rate of 16 kHz, as well as their corresponding clean utterances. The test set (target domain) contains $N_T=824$ noisy utterances with 5 types of unseen noise in SNR levels (2.5, 7.5, 12.5, and 17.5 dB).\par

\noindent \textbf{TIMIT}. To evaluate the proposed method in serious domain mismatch and low SNR conditions, we use clean utterances from TIMIT to customize the source and target samples. The training set contains 576 utterances, contributed by 48 male and 24 female speakers from 8 dialect regions. These clean utterances are mixed with 5 stationary noise types (\textit{car}, \textit{engine}, \textit{pink}, \textit{wind}, and \textit{cabin}) at 4 SNR levels (-6, 0,  6, and 12 dB), amounting to 11520 noisy utterances, to be the paired data from source domain $\{X_i^S,Y_i^S\}_{i=1}^{N_S}$ with ${N_S}= 11520$. For the target domain, we employ the 192 clean utterances from test set, which were subsequently mingled with one of the 2 non-stationary noise types (\textit{helicopter}, \textit{baby-cry}) under 5 SNRs (-6, -3, 0, 3 and 6 dB) as target input $\{ x_i^T\}_{i=1}^{N_t}$ with $N_T=576$. The choice of noise types for the source and target domain was to let the learning algorithms adapt from distinguished environments in the real world.\par

\subsection{Training and Evaluation}
\noindent \textbf{Configuration}.
For UNA-GAN, the magnitudes are all cut into segments with the dimension of $129\times 128$. The ResNet block is repeated for 9 times, so the first and fifth blocks are selected for contrastive learning. The self-attention layers are repeated 3 times. In Eq.~(\ref{loss_total}), $\alpha$ and $\beta$ are all set as 1. For SE model, the TCN blocks are stacked 4 times. The initial learning rates for UNA-GAN and SE model are respectively 0.002 and 0.001, and both networks are optimized by the Adam algorithm~\cite{kingma2014adam}.\par
\noindent \textbf{Metric}. We employ perceptual evaluation of speech quality (PESQ)~\cite{rix2001perceptual} as main metric to evaluate the performance of SE model. Furthermore, we report prediction of the signal distortion (CSIG), prediction of the background intrusiveness (CBAK), and prediction of the overall speech quality (COVL)~\cite{hu2007evaluation} for comparison with other works. For all metrics, higher scores mean better performance.

\subsection{Reference Baseline}
To evaluate the effectiveness of the proposed UNA-GAN, we built 4 baselines for comparison. It is worthy noted that NAT-SE and DAT require the domain label of noise during training, thus is considered as \emph{weakly supervised} method. \par
\noindent \textbf{Vanilla-SE} trains the SE model (right of Fig.~\ref{f1}) only using source domain data without any adaptation.\par
\noindent \textbf{NAT-SE}~\cite{hou2021learning} learns disentangled features by a further discriminator module, which is trained on the VoiceBank dataset. \par
\noindent \textbf{DAT}~\cite{liao2018noise} introduces the domain adversarial training that utilizes a domain classifier on TIMIT dataset. \par
\noindent \textbf{Upper-bound} trains the SE model using source domain and then adapts it using labeled target domain data, which can be viewed as upper-bound performance for noise adaptation. \par

\section{Result and Analysis}

\subsection{Data requirement of target domain}
To demonstrate the data efficiency of UNA-GAN, we first evaluate the performance on Voicebank-DEMAND with different data amounts of target domain, which is often limited in practical conditions. The main results for each unseen noise category are shown in Table~\ref{t1}, where $N_t$ denotes the number of utterances from target domain that randomly selected from $\{x_i\}_{t=1}^{N_T}$. The maximum of $N_t$ is 160 (6.8 minutes). We observe that UNA-GAN method achieves the noise adaptation when only 1.7 minutes of target domain data is available, which demonstrates the data efficiency. Furthermore, the PESQ performance obviously benefits from the increase in data amount for all noise types.  

\begin{table}[t]
\caption{PESQ results of UNA-GAN with different amounts of target domain utterances $N_t$.}
\vspace{-0.1cm}
\centering
\resizebox{0.48\textwidth}{!}{%
\begin{tabular}{c | c | c| c c c c c}
\toprule[1.2pt]
\multirow{2}{*}{ID} & \multirow{2}{*}{System}  & \multirow{2}{*}{$N_t$}    & \multicolumn{5}{c}{Noise type}  \vspace{0.02cm} \\ 
   &                    &          & \emph{Cafe} & \emph{Living} & \emph{Office} & \emph{Psquare} & \emph{Bus}   \\ \midrule[1.2pt]
1 &  Unprocessed                  &     -    &1.49 &  1.61    &  2.53   &   1.74  &  2.48 \\ \midrule
2 &  Vanilla-SE  &   0    &   2.29   &   2.56   &   3.01 & 2.56  & 3.11  \\ \midrule

3 &  \multirow{3}{*}{UNA-GAN}     &   40     & 2.34 &  2.65 & 3.13 & 2.61 & 3.22\\ 
4 &                               &   80     & 2.38 &  2.71 & 3.18 & 2.65 & 3.30\\
5 &                               &   160    & 2.40 &  2.73 & 3.22 & 2.69 & 3.31       \\ \bottomrule[1.2pt]
\end{tabular}}
\label{t1}
\end{table}

\subsection{Result on Voicebank-DEMAND}
We then report the results on Voicebank-DEMAND dataset that adapt to 5 unseen noises using single SE model. System 3$\sim$5 employs the same TCN-based SE model, while NAT-SE requires domain labels of noises for adversarial training. We observe that the proposed UNA-GAN surpasses other baselines by a large margin in terms of all metrics and achieves comparable performance with the upper-bound system that is finetuned by labeled test set.   

\begin{table}[t]
\caption{Result on Voicebank-DEMAND. ``D.L." denotes whether system requires domain labels during training.}
\vspace{-0.1cm}
\centering
\resizebox{0.48\textwidth}{!}{%
\begin{tabular}{c | c | c| c c c c c}
\toprule[1.2pt]
ID &  System                 &   D.L.  & PESQ & CSIG & CBAK & COVL &    \\ \midrule[1.2pt]
1  &  Unprocessed            &     -    & 1.97 & 3.35 & 2.44 & 2.63 &   \\ \midrule
2  &  Vanilla-SE             &    \xmark   & 2.67  &  3.93 &  3.29 &   3.30   \\ 
3  &   NAT-SE                &    \cmark   & 2.72  &  3.99  &  3.47 &  3.36  \\
4  &   UNA-GAN               &    \xmark   & 2.91  &  4.05 & 3.54 &   3.43    \\ \midrule
5  &   Upper-bound           &    \xmark   & 2.95  &  4.11    & 3.59   & 3.52     \\ \bottomrule[1.2pt]
\end{tabular}}
\label{t1}
\end{table}

\subsection{Result on TIMIT}
In the last experiment, we explore the effect of UNA-GAN in face of larger domain mismatch and lower SNR levels. The PESQ results are shown in Table~\ref{table3}. It is noted that the UNA-GAN only leverages 0dB target domain data ($N_t$=192) for adaptation and testing in all SNR levels.\par 
We observe that the Vanilla-SE losses effectiveness of enhancement when directly test on unseen noises, especially in low-SNR settings. Despite only simulating 0 dB data, the PESQ performance of UNA-GAN increases obviously in all SNR conditions. Furthermore, it respectively surpasses DAT baseline by 15.2\% and 6.6\% for \textit{helicopter} and \textit{baby-cry} noises on average.      

\begin{table}[]
\caption{PESQ results on TIMIT dataset with different SNRs. ``Avg" denotes the average of all SNR levels.}
\vspace{-0.1cm}
\resizebox{0.48\textwidth}{!}{\begin{tabular}{c|c|c|cccccl}
\toprule[1.2pt]
\multirow{2}{*}{ID} & \multirow{2}{*}{System}     &  \multirow{2}{*}{D.L.}     &   \multicolumn{6}{c}{Noise level, SNR =}                  \\
                  &           &  &  -6  &  -3         & 0         & 3    & 6  & Avg.        \\ \midrule[1.2pt] 
\multicolumn{9}{c}{\cellcolor[HTML]{E0E0E0}\emph{Noise type: Helicopter}} \\ 
1&Unprocessed &  \xmark & 1.05  & 1.07 & 1.10 & 1.16 & 1.26 & 1.13 \ \textcolor{red}{+0\%} \\
2&Vanilla-SE  &  \xmark & 1.06  & 1.09 & 1.18 & 1.28 & 1.44 & 1.21 \ \textcolor{red}{+7.07\%}  \\
3&DAT         &  \cmark & 1.15  & 1.20 & 1.27 & 1.52 & 1.78 & 1.38 \ \textcolor{red}{+22.1\%}    \\ 
4&UNA-GAN     &  \xmark & 1.21  & 1.32 & 1.55 & 1.78 & 2.08 & 1.59 \ \textcolor{red}{+40.7\%} \\ \midrule
\multicolumn{9}{c}{\cellcolor[HTML]{E0E0E0}\emph{Noise type: Baby-cry}} \\ 
5&Unprocessed &  \xmark & 1.06  & 1.09  & 1.13 & 1.18 & 1.27  & 1.15 \ \textcolor{red}{+0\%}  \\
6&Vanilla-SE  &  \xmark & 1.07  & 1.09  & 1.15 & 1.22 & 1.41  & 1.19 \ \textcolor{red}{+3.48\%}   \\
7&DAT         &  \cmark & 1.31  & 1.50  & 1.62 & 1.80 & 2.13  & 1.67 \ \textcolor{red}{+45.2\%}  \\ 
8&UNA-GAN     &  \xmark & 1.40  & 1.57  & 1.74 & 1.96 & 2.21  & 1.78 \ \textcolor{red}{+54.8\%} \\ \bottomrule[1.2pt]
\end{tabular}}
\label{table3}
\end{table}

\begin{figure}[h]
\centering
\includegraphics[width=0.45\textwidth]{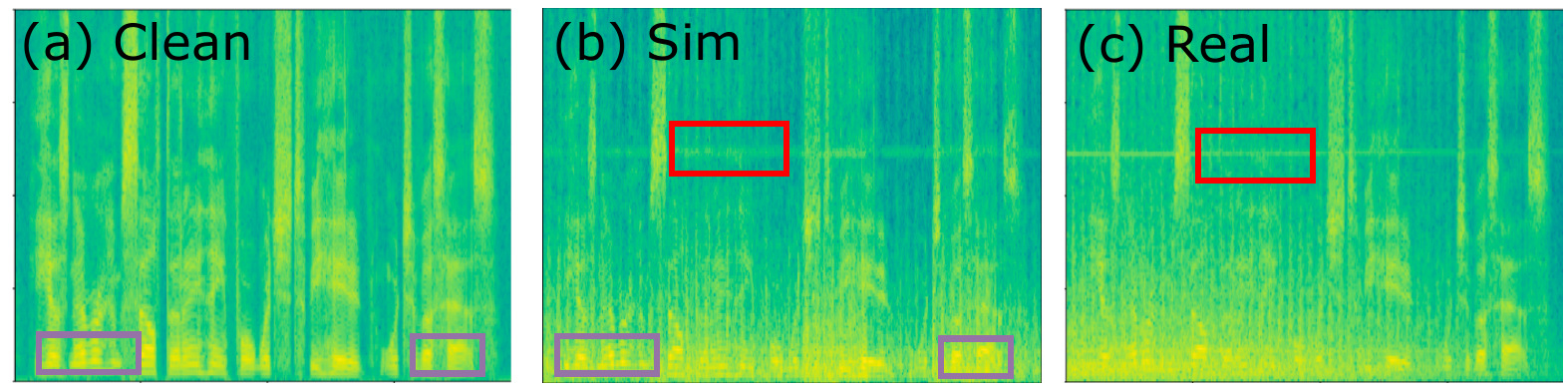}

\caption{The magnitude of a sample (SI1039.wav). (a) is the clean magnitude, (b) is the simulated magnitude by UNA-GAN, and (c) is the noisy magnitude from target domain (ground-truth), which is a 0 dB \textit{helicopter} noisy utterance.}
\vspace{-0.1in}
\label{f2}
\end{figure}

To visualize the effect of UNA-GAN, we sample and draw the clean, simulated, and real magnitudes in Fig~\ref{f2}. It is observed that the simulated magnitude has learned the similar distribution of \textit{helicopter} noise in target domain. Specifically, the simulated \textit{helicopter} magnitude appears same horizontal stripe (red box) and vertical bands with real magnitude from target domain. Meanwhile, we also observe that some invalid speech information of speaker has been retrained in simulated magnitude (purple boxes), which is contributed by the multi-layer contrastive learning.  
\vspace{-0.1in}
\vspace{-0.1in}

\section{Conclusion}
We address the unsupervised noise adaptation issue in speech enhancement. The proposed UNA-GAN method learns a clean-to-noisy transformation by several minutes of unpaired data and then adapts SE model to target noise by simulated data. Experimental results show that UNA-GAN effectively increase SE performance in terms of evaluation metrics, even in face of large domain mismatch and low-SNR conditions.

\vfill\pagebreak


\bibliographystyle{IEEEtran}
\bibliography{refs}

\end{document}